# Too big to see: exploring proxies of structure in a real large-scale university-industry cooperation network

Yuri Campbell

Fraunhofer Center for International Management and Knowledge Economy (IMW), Leipzig, Germany

**Abstract:**

We investigate static and dynamic topologies of a 2-mode real world university-industry cooperation network. Due to its large size and complex structure, we choose to use statistical proxies for this goal. Among the findings, we shall call attention to the rank-size distribution of the firm node degrees with power law's signature log-linear behavior. Which invokes hints of a complex network architecture, as a counterpoint to the random case. We compare furthermore the rank-size distributions of both modes with other real-world 2-mode large-scale networks and draw parallels for their causes. Moreover, we investigate structural change in the network by computing Robins-Alexander clustering coefficients in a rolling window fashion in order to capture topological change in the network temporal evolution. Findings suggest that network stability is achieved after a short transition phase of high clustering and cross-correlation.

1. Introduction and aim

Open forms of collaboration can contribute in various forms in order to enhance the process of innovation in enterprises. Alone the open exchange of ideas can help in identifying new business opportunities, uncovered market shares or needs. Collaborative innovation can involve distinct layers of the production chain and have different goals. It might involve customers, competitors and suppliers as well. Because it may take place in all aspects of the business cycle, from the procurement and supplier collaboration all the way to distribution and commercialization. Here though, we focus on the research and development of new products, services and technologies. The object of our study is collaborative innovation in research and development (R&D) between firms and public research organizations (PROs).

Collaborative innovation is a social phenomenon itself (Gloor, 2006). Therefore, it is not surprising that it is a process embedded in a network structure. Due to a growing body of empirical research, it is continuously well accepted that innovation by individuals or large-scale units, as teams or institutions, is influenced by the relationships they share, in other words, by the local structure they experience in the embedding underlying (social) network (Guan & Liu, 2016). For example, empirical work by Guan and Zhao (2013) suggests that when researchers participate in networks with small-world property, they would have more chance to acquire fresh and unfamiliar information easily. Which supports the idea of the strength of weak ties in research of innovation (Gretzinger, Hinz, & Matiaske, 2011).

Often, research in this direction focus on patent data, and therefore it is natural to choose a testbed domain as biopharmaceuticals (Guan & Zhao, 2013) or nano-energy (Guan & Liu, 2016). Which is *per se* important as a hypothesis-testing model, but admittedly very restrictive. This is problematic for two reasons. First, the availability of R&D and patent data is dependent on the sector. Second, the majority of firms, which are small-and-medium sized, do not report their R&D activities and do not have the resources or capabilities to file patents. In summary, a lot of structure is lost in focusing on sector and patent-data. Additionally, most of these studies miss a crucial structural property of collaborative innovation in R&D between firms and PROs, the fact that the underlying network is bimodal.

A bipartite network, or in other words a 2-mode/bimodal network (Latapy, Magnien, & Del Vecchio, 2008), is a network in which the participants can be divided in two classes, for example firms and PROs, and interactions happen just from one class to the other. While small networks are in somewhat trivial for the machinery in our visual cortex in charge of pattern recognition, for large-scale networks (uni- and bimodal) we rely on statistics and computing power to grasp their structure. Unimodal large-scale real-world networks share nontrivial properties that differ from most networks of the same size (Albert & Barabási, 2002). Properties which not just social networks, but also technological, biological and linguistic ones share, efficiently reviewed by Barabási (2016) and references therein. The most notable among those and almost universal is the peculiar degree distribution, which follows a power law. In addition, characteristics from cluster index and community formation to redundancy differ in general from a random case, but are differently characterized depending on the network in case. The same holds for bimodal large-scale real world networks (Latapy et al., 2008). However, in general, structural properties there are richer, with fewer universal one-fits-all properties among all classes of bimodal real networks. This motivates the study and comparison of real bimodal large-scale networks in a case-by-case basis.

Our goal is to explore a real large-scale collaborative innovation network in R&D, to which data we were granted access, and to compare its structural properties and temporal dynamics, using statistical proxies common in large-scale network analysis, against other examples of 2-mode real world networks.

## 2. Research methodology

We study the collaborative innovation network in R&D between the research units of the largest organization for applied research in Europe, the Fraunhofer Society, and its R&D customers. The Fraunhofer Society documents all the projects with industry-partners. As these are usually contracted research projects, which alone for financial and controlling reasons must be registered and reported. These intern project data are the object of study of our analysis. Different from existing approaches, which mostly use patent data to this end, the used dataset covers much more of the science spectrum in applied research. This is due to the diverse research portfolio of the organization in study (Comin, Licht, Pellens, & Schubert, 2019; Fraunhofer, 2017), which includes life sciences, materials and microelectronics, to name a few.

In practice, our dataset contains roughly a hundred thousand collaboration projects and subprojects in R&D dated from the year 1992 until 2018 carried out between 74 Fraunhofer independent research units and companies, excluding classified projects (e.g. with the defense industry). Due to the nature of the data provided, the division into two groups, PROs and firms, is quite clear. This together with their collaboration activities, registered as cooperation projects, define clearly a bipartite collaborative innovation network. Which is moreover *per se* cross-domain, another advantage of analyzing internal data.

From the original data set, the construction of a bipartite network is straightforward. For this we use the software package NetworkX (Hagberg, Swart, & Chult, 2008) and its procedures therein, for simplicity. With the data loaded in a bipartite network model, simple statistics, like the degree distribution for both modes can be computed. For this, if a firm and a PRO ever cooperated, they will be linked with an edge.

Hence, for the degree distribution, we were interested in the general structure of the collaborative network, independently of the time the collaboration occurred.

Now, in a second analysis step, in order to access the structural evolution dynamics of the network, we opted to look at the time evolution of the Robins-Alexander clustering coefficient (Robins & Alexander, 2004) using a variable-length sliding window based on the year of beginning of every cooperation project. The Robins-Alexander clustering coefficient is a natural generalization for bipartite networks from the standard clustering coefficient in one-mode networks. The later is the probability when three nodes are connected in a chain (by two edges), that they form a triangle. While the former is the probability when three nodes are connected in a chain, that they form a square (Latapy et al., 2008). In our case, the Robins-Alexander clustering coefficient translates to the probability when firm *A* and *B* collaborated with PRO *X*, and firm *B* with PRO *Y*, that firm *A* also collaborated with PRO *Y* (Figure 1). This is intuitive because in our dataset there are no triangles, that is, no collaboration between PROs. Moreover, it is clear that the clustering coefficients take into account the whole network.

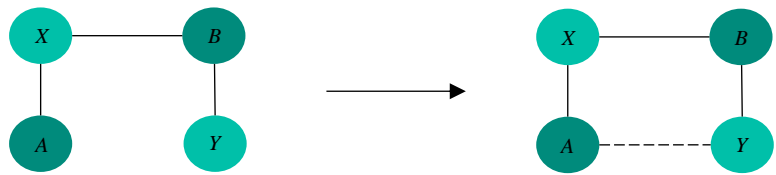

*Figure 1. Robins-Alexander intuition: clustering coefficient based on squares.*

For this second analysis, two edges in the network were connected if a project happened in the time frame of the variable-length sliding window. The Robins-Alexander clustering coefficients were calculated for every possible combination of window length and initial year from 1992 to 2018. Naturally, for every window, there is a unique network topology.

3. **Results & discussion**

In Figure 2, we show the rank-size degree distribution for firms on the right side and the research units of the Fraunhofer Society on the left side. The company node degree distribution (right side) strongly hints to a family of power law distributions, due to its log linear pattern. As well as in other real-world 2-mode networks, as the distribution of degrees for co-authors in a bipartite network between authors and papers on *arXiv*, or the distribution of degrees for co-appearance of words in a network between sentences and word from a large text corpus (Latapy et al., 2008). Another interesting fact, the rank-size degree distribution for the Fraunhofer research units follow a similar degree distribution of both papers and sentences, with a concave pattern slightly right-skewed. On the original study, authors point this to the fact that the nodes of the mode in question can saturate. For example, the number of words is in a sentence limited. Likewise, papers have a limited number of authors. In our case, this may hold, but with a couple of caveats. Both capacities of firms and research units are limited, clearly. However, the number of research units is up to three orders of magnitude inferior as the number of nodes from the opposite mode. That is, although the capacity of firms is limited, it is almost never saturated.

Now, instead of static degree distribution statistics, Figure 3 proxies the temporal evolution of the topology of the collaborative network. The plot shows the Robins-Alexander clustering coefficient for every possible combination of sliding window. On the right side, we include the year 2008 and, on the right side, we exclude it. From the plot on the left, one easily sees that the network undergoes a transition period of high clustering in the year 2008. The values are comparably so inflated that pre- and post-2008 look alike, with low coefficients. Hence, we remove from the plot only this year and replot the same statistics; this is shown on the right side. Now we observe that the two eras, post- and pre-2008, also have distinct clustering coefficient patterns. Moreover, the phase post-2008 has on average higher clustering coefficient than the phase pre-2008. All the above-observed hint to a phase transition in the collaborative network. Which is marked with a (short) transition period with high correlation among the nodes.

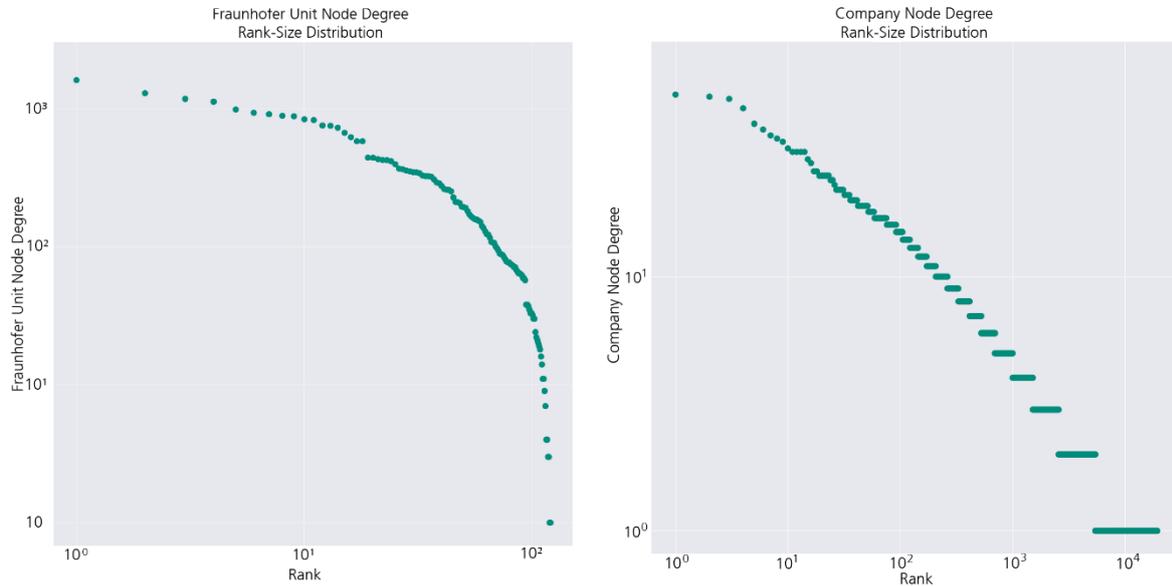

*Figure 2. Rank size distributions for each of the two modes of the collaborative innovation network.*

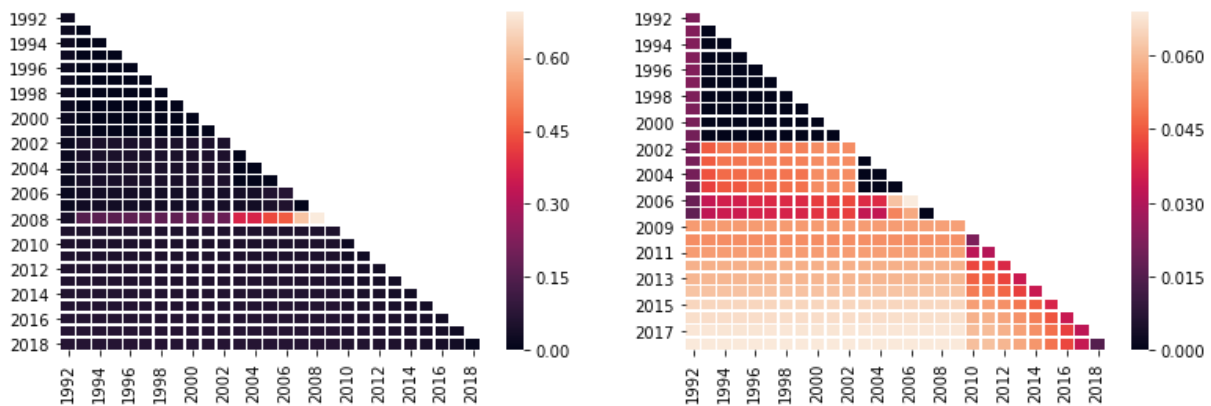

*Figure 3. Robins-Alexander clustering coefficients for all network topologies. On the left with the year 2008, on the right without. The start of the window is the beginning of the year on the X-axis, and the end is the end of the year on the Y-axis.*

## 4. Final Remarks

The shown preliminary results about the analysis of a real large-scale collaborative innovation network in R&D point to an overall complex structure, with nontrivial topological evolution behavior. Although such results shall not be as surprising for a large organization. It is nevertheless striking that some of its statistical properties correspond so well with well-known large-scale 2-mode networks. Moreover, it is remarkable that the Robins-Alexander clustering coefficient could capture a phase transition in the collaborative innovation network, from a network of more isolated communities to a network of more complex structure. This marks a first step in order to understand how this collaborative innovation ecosystem works. Such that, in the future we can optimize knowledge transfer activities between both modes of the network and increase innovation output.